# Deep caller for ocean acoustic releases

by Hans van Haren[*], Martin Laan, Sander Asjes and Bas Denissen

Royal Netherlands Institute for Sea Research (NIOZ) and Utrecht University, P.O. Box 59,
1790  AB Den Burg, the Netherlands.
*Corresponding author e-mail: hans.van.haren@nioz.nl

ABSTRACT

We relate about the custom-made modification of a Benthos deep-ocean acoustic release into a 'deep caller', a hydrophone for calling ocean acoustic releases that cannot be reached from a standard deck-unit. The self-contained deep caller can be lowered down to 12 km on any non-conducting winch cable. It may prove useful to retrieve sub-surface instrumentation like a seafloor-lander hidden behind large rocks or in a narrow canyon, or moorings in very deep topographic depressions. We used it to retrieve a 7 km long mooring from 10,910 m depth in the Challenger Deep, Mariana Trench.

**1. Introduction**

Ocean-researchers who regularly deploy stand-alone sub-surface instrumentation in the deep-sea may have experienced difficulties in retrieving their precious scientific materials. Normally, such instrumented lines or landers are equipped with one or two devices to release one or more weights acoustically from a deck-unit on board a ship. A standard deck-unit is attached to a hydrophone via a cable of a few tens of meters length only. When sufficient buoyancy materials are applied the instrumentation will surface after releasing its 'anchor' weight(s). However, acoustic releases sometimes fail for various environmental reasons other than electronic failure when meticulous maintenance is performed. Environmental reasons include: A rocky and/or canyon-like seafloor, so that acoustic signals may become reflected leading to loss from their precise frequency modulation band, a strong bubble or pycnocline layer that may divert the acoustic signal, and (too) great distances between the deck-unit hydrophone and the acoustic releases near the seafloor. In these cases it may be helpful to have the ability to lower a hydrophone nearer to the deep-sea acoustic releases. This short paper narrates the design, construction and successful use of such deep acoustic release 'caller', for Teledyne Benthos releases.

**2. Design of the deep caller**

In order to reach deep or hidden acoustic releases of sub-surface ocean moorings and landers, a custom-made self-contained hydrophone was designed and built at NIOZ, the Royal Netherlands Institute for Sea Research. As NIOZ commonly operates two releases in tandem on heavily instrumented mooring lines and landers, the requirement was to have the possibility to call two release codes alternately. Its depth rating should be full ocean depth (>11 km).

*a. Remodeling an acoustic release into a deep caller*

A second-hand Benthos deep-ocean 865-A was gender-changed from being an acoustic release to become a self-contained deep hydrophone caller, to be lowered into the ocean by a ship's or helicopter's winch cable. Its depth rating is 12 km. The entire final stage of electronics was disconnected from the high-voltage transformer. The transformer was connected to a newly designed push-pull final stage of electronics. This processor-directed final stage controls in- and out-band frequencies as in a standard deck-unit.

More acoustic transmission power was needed. Therefore, the power supply was modified by adding two times 9 chargeable NIMh HR-4/3FAU cells of 4.5 Ah. Charging the battery packs takes about 14 h. These extended battery packs last at least 5 h when programmed to ping release codes for 10 s every 120 s. At a typical cable lowering speed between 0.5 and 1 m/s, any seafloor can be reached pinging, even in the deepest spot on Earth. After programming, the deep caller is switched externally via the screw-switch like in Benthos acoustic release activation.

*b. Software program*

A custom-made program allows the calling of two different Benthos acoustic releases, at selectable time intervals. The program demands the main release 'in-band' frequency, its narrow modulation bandwidth and its out-band frequency. When two different release codes are programmed, these will be transmitted alternately, with a certain delay in between. Transmission time-length and delay time-length can be programmed independently. To prevent the caller from overheating, it is recommended to restrict transmission to less than 20 s per call, and to allow at least 60 s of quiescence (delay) between transmission.

**3. Using a deep caller at sea**

*a. Seatest*

As a first test, we programmed the caller with two release codes, allowing for 10 s of transmission and 120 s delay between two transmissions. Two different acoustic releases were attached to the Rosette-frame of a shipborne Conductivity Temperature Depth CTD. The CTD was lowered to 2 km below the ship in the open East-Atlantic Ocean. Using a separate winch, the switched-on deep caller was lowered to 100 m below the ship to be certainly deeper than the diurnal pycnocline.

The 120 s delay of quiescence gave ample time to listen to the releases' responses via a standard Benthos deck-unit hydrophone. One release responded immediately upon the very first deep call under water. The second release did not respond after repeated calls, presumably because of signal transmission via the CTD-conducting cable.

*b. Challenger Deep release*

For a study on internal wave turbulence in hadal zones we designed 300 high-resolution temperature sensors rated to 1400 Bar and which were moored for nearly three years at about 10,910 m water depth in the Challenger Deep, Mariana Trench, close to the deepest point on Earth (van Haren et al. 2017). The 7 km long mooring was equipped with a pair of Benthos deep-ocean/865-A releases, at 6 m above the anchor-weight. The releases could not be reached with a standard Benthos DS8000 deck-unit. Steep slopes are observed on the accretionary prism for the last 10 km before reaching the deformation front, with water depth increasing from 8300 m to 10,925 m. The Challenger Deep is long, but quite narrow: 120 km wide when measured at 5000 m water depth in the study area, less than 1 km wide for the lower 25 m above the seafloor.

A deep caller was attached to R/V Sally Ride's steel cable just above a 50 kg weight, holding the caller upside down and using a cable stocking for security. A standard Benthos deck-unit with 10 m of hydrophone cable was used to listen to the releases' response.

After one-and-a-half hours of lowering not a single response was received on the deck-unit from the deep releases when the mooring top-buoy surfaced, as was noticed via signals from the radio beacon and the Iridium satellite beacon mounted on the top-buoy. During the operation, the R/V Sally Ride was held in position via dynamic positioning. In hindsight, acoustic noise due to turbulence generated by the impellors for dynamic positioning may have masked reception of the weak return signal at the deck-unit. This probable adverse effect of dynamic positioning on the reception of response signals from acoustic releases was later confirmed by British colleagues.

Reconstructing after the successful mooring recovery, the deep caller had managed to contact the near-bottom releases after about 15 to 20 minutes of lowering, when it was around 1000 m depth or at a distance of approximately 10 km from the mooring releases. This is calculated from the time for the top-buoy to surface from 4 km, which takes about one hour, while the first Iridium beacon messages arrive about 15 min after surfacing. Both deep-ocean Benthos mooring releases had worked perfectly, some 40 months after being switched on and under more than 1100 Bar of ambient pressure.

**4. Concluding remarks**

While we successfully released our mooring from the deepest point on Earth, the self-contained deep caller may be helpful especially for small bottom-landers that easily hide in rugged terrain such as in canyons. It may also be helpful under conditions when a lander is toppled over, when landed on its side. Imaginary use is from a helicopter or from a free-floating device, when a 'hand-held' hydrophone is not easily operated from a deck-unit. The same may be said for releasing from a ship under strong surface current conditions, when a light-weight deck-unit hydrophone is difficult to direct vertically, as occurred during our first retrieval cruise from the R/V Sonne in the Challenger Deep. The (weighted) deep-caller is much easier to direct vertically than a standard hydrophone.

We described a deep-caller for Benthos releases only. Although electronics are different, principles are portable to other release systems.

*Acknowledgements.* We thank the masters and crews of the R/V Sonne and R/V Sally Ride for the pleasant cooperation during the operations at sea.